\documentclass[iop,apj]{emulateapj}

\newcommand{\etal}{et al.}

\newcommand{\sigmagas}{$\Sigma_{\mathrm{gas}}$}
\newcommand{\sigmastar}{$\dot{\Sigma}_{\ast}$}

\slugcomment{Accepted to The Astrophysical Journal}

\shorttitle{The Star Formation Laws of Eddington-Limited Star-Forming Disks}
\shortauthors{Ballantyne \etal}

\usepackage{times}

\begin{document} 

\title{The Star Formation Laws of Eddington-Limited Star-Forming Disks}


\author{D. R. Ballantyne, J. N. Armour and J. Indergaard}
\affil{Center for Relativistic Astrophysics, School of Physics,
  Georgia Institute of Technology, Atlanta, GA 30332;
  david.ballantyne@physics.gatech.edu}

\begin{abstract}
Two important avenues into understanding the formation and evolution of
galaxies are the Kennicutt-Schmidt (KS) and Elmegreen-Silk (ES)
laws. These relations connect the surface densities of gas and star
formation (\sigmagas\ and \sigmastar, respectively) in a galaxy. To elucidate the KS and ES
laws for disks where \sigmagas$\ga 10^4$~M$_{\odot}$~pc$^{-2}$, we
compute 132 Eddington-limited star-forming disk models with radii spanning tens to
hundreds of parsecs. The theoretically expected slopes ($\approx 1$ for
the KS law and $\approx 0.5$ for the ES relation) are relatively robust to spatial
averaging over the disks. However, the star formation laws exhibit a
strong dependence on opacity that separates the models by the
dust-to-gas ratio that may lead
to the appearance of a erroneously large slope. The total infrared luminosity ($L_{\mathrm{TIR}}$)
and multiple carbon monoxide (CO) line intensities were computed for each
model. While $L_{\mathrm{TIR}}$ can yield an estimate of the average
\sigmastar\ that is correct to within a factor of 2, the
velocity-integrated CO line
intensity is a poor proxy for the average \sigmagas\ for these
warm and dense disks, making the CO conversion
factor ($\alpha_{\mathrm{CO}}$) all but useless. Thus, observationally derived KS and ES
laws at these values of \sigmagas\ that uses any transition of CO will
provide a poor measurement of the underlying star formation
relation. Studies of the star formation laws of Eddington-limited
disks will require a high-$J$ transition of a high density molecular
tracer, as well as a sample of galaxies with known metallicity estimates.
\end{abstract}


\section{Introduction}
\label{sect:intro}
Galaxies can experience a wide variety of star formation phenomena, from the
relative calm of star-forming regions beaded along a spiral arm,
to the fury of a nuclear starburst burning through a trillion
Suns. Yet, remarkably, the rate of star formation in such wildly
disparate environments can be simply related to the density of
gas in the star-forming region, $\dot{\Sigma}_{\ast} \propto
\Sigma_{\mathrm{gas}}^N$, where \sigmastar\ and \sigmagas, are the star
formation rate (SFR) and total gas (i.e., atomic plus molecular)
surface densities. This relation, known as the Kennicutt-Schmidt (KS) law,
has been observed to hold over nearly four decades in \sigmagas\ with
$N\approx 1.4$ \citep[e.g.,][]{sch59,kenn98,kenn98b,yao03,ke12}, and is ultimately
related to the efficiency of star formation and its connection to both local and
global timescales
\citep[e.g.,][]{elm97,silk97,kenn98b,elm02,km05,krum09}.  The
super-linear slope indicates that the star formation efficiency (i.e., the
ratio of stellar mass formed to the gas mass in the star-forming
region) increases
at high densities. Measuring and understanding the KS law over
a wide range of star-forming environments is crucial to elucidating
comprehensive theories of star and galaxy formation (see discussion by
\citealt{ke12}).

Most of the star formation in the universe occurred long ago at $z > 1$
when the majority of massive galaxies were being assembled
\citep[e.g.,][]{hb06,pg08}. With the advent of new sensitive millimeter arrays and
detectors, the last decade has seen a significant advance in the study
of the KS law at high redshifts (e.g., \citealt{daddi10,gen10,ivi11,fre13};
see \citet{svb05} and \citet{cw13} for reviews). Interestingly, these galaxies, which have both much larger
star formation rates and gas densities than local objects, seem to
follow a similar KS slope, but are vertically offset from the local
relation \citep{daddi10,gen10} perhaps implying a fundamental difference in the stellar initial mass function
(IMF) in the two regimes. However, many of these results rely on the assumption that the
troublesome conversion factor
$\alpha_{\mathrm{CO}}=\Sigma_{\mathrm{gas}}/I_{\mathrm{CO}}$ needed
to convert from the observed CO velocity-integrated intensity, $I_{\mathrm{CO}}$, to
\sigmagas\ is bimodal with one value ($\alpha_{\mathrm{CO}} \approx 4$)
for galaxies with Milky Way-like SFRs, and another value
($\alpha_{\mathrm{CO}} \approx 0.8$) for galaxies with much larger
SFRs (i.e., high-$z$ galaxies). When \citet{nar12} recently recomputed the KS
law with a value of $\alpha_{\mathrm{CO}}$ that varied continuously
with $I_{\mathrm{CO}}$, the relation was no longer bifurcated between
the high and low-$z$ galaxies and could be fit over a wide range in
\sigmagas\ with $N\approx 2$ (see also \citealt{os11}). 

Alternatively, it has been argued \citep{elm97,silk97,daddi10,gen10} that a more fundamental star
formation law is $\dot{\Sigma}_{\ast} \propto
(\Sigma_{\mathrm{gas}}\Omega)^{n}$, where $\Omega$ is the orbital angular frequency in the star-forming
region, usually estimated at the observed outer
radius of a galaxy. Both high and low-$z$ galaxies seem to follow this
Elmegreen-Silk (ES) relation with $n\approx 1$, independent of the exact assumptions on
$\alpha_{\mathrm{CO}}$ (\citealt{daddi10,gen10}; although \citet{nar12} found that the scatter in
the ES relation was reduced by using the continuously varying description of
$\alpha_{\mathrm{CO}}$). In this formulation, a larger star-forming
efficiency is found in
rapidly star-forming galaxies because of occurring in more compact
regions with shorter dynamical times.

Analysis and interpretation of the star formation laws at high
redshift are hampered by gaps in the theory of star formation in
galaxies at these redshifts. The interstellar medium of these galaxies
are expected to be predominately molecular
\citep[e.g.,][]{sol97,tac10,daddi10a,mag12} have gas fractions of several tens of percent \citep[e.g.,][]{geach11,swin12,nbd12} with significant turbulence and
large pressures \citep[e.g.,][]{ds98,genz08,wis11}. Simulating this environment from first
principles is computationally prohibitive, so analytical models have
been developed that encompass many aspects of the physics in a
well-defined region of parameter space \citep[e.g.,][]{tqm05,os11}. In addition, as eluded
to above, there may be severe uncertainties in relating the observed quantities
(usually, infrared (IR) luminosity and CO intensity) to the physical
parameters \sigmastar\ and \sigmagas\ \citep[e.g.,][]{svb05,feld12,nar12}. Insofar as the IR luminosity
captures the bolometric output of embedded star-forming regions, there
are well known calibrations between the observed IR luminosity and SFR
\citep[e.g.,][]{kenn98b,rieke09,calz10,murph11}.  More
problematic is how to convert from the line intensity of a molecule
that comprises only a tiny fraction of the interstellar gas to an
estimate of the entire gas mass. Although the CO lines are most
commonly used due to its high abundance, it has been noted that
because of their relatively low critical densities and high optical
depth, they are likely very poor tracers of high density gas \citep[e.g.,][]{kt07,nar08}. Indeed, KS laws derived from a high density tracer such as HCN
have produced a simple linear relation between \sigmastar\ and
\sigmagas\ for both high- and low-$z$ galaxies \citep[e.g.,][]{gs04a,gs04b}, although the conversion factor $\alpha_{\mathrm{HCN}}$ is
just as uncertain as the one for CO (see the discussion by \citealt{gb12}). Taken together, the combined
uncertainties in both the existing theoretical framework and the
observational conversions severely restrict the use of the KS or ES
laws.

The most extreme star-forming regions are those
with $\Sigma_{\mathrm{gas}} \ga
10^4$~M$_{\odot}$~pc$^{-2}$. In this situation the opacity of the dense and dusty gas is
so large that the ISM becomes optically thick in the IR, radiation
pressure dominates over the turbulent pressure, and the disk is said
to be Eddington-limited. Expected to occur in the cores (i.e., central
few hundred pcs) of rapidly
star-forming disks, these intense star-forming regions are difficult
to investigate observationally, although evidence for such high values
of \sigmagas\ have been recently inferred at the centers of some
$z < 1$ post-starburst galaxies \citep{ds12} and
star-forming systems \citep{gb12}. Their location also implies that
these Eddington-limited star-forming disks may act as a `bridge'
between the black hole environment and the host galaxy, ferrying fuel
through the disk toward the central black hole and potentially driving
nuclear activity \citep{tqm05,ball08}. Thus, predictions of the observational
signatures of Eddington-limited star-forming regions are needed to
guide future observational studies of the centers of both active and
inactive galaxies at all redshifts.  \citet{os11} analytically investigated star
formation in environments with $10^2 \la \Sigma_{\mathrm{gas}} \la
10^4$~M$_{\odot}$~pc$^{-2}$ when vertical pressure support is provided by
supernova driven turbulence, and predicted that the KS law should have
$N=2$ (see also \citealt{tqm05}),
in decent agreement with the high-$z$ observations if one assumes that
$\alpha_{\mathrm{CO}}$ is inversely correlated with
$I_{\mathrm{CO}}$. In the Eddington-limited case the local KS law is
expected to flatten to $N=1$ \citep{tqm05,os11}, but there have been no predictions for the
ES law nor work on how clearly these laws will translate for an observation that
encompasses an entire $>100$~pc star-forming disk and contains
a significantly variable SFR and \sigmagas. In this paper we use the \citet{tqm05} model
of an Eddington-limited disk to self-consistently
calculate the radial structure of over 100 nuclear star-forming disks,
and, through the use of CO radiative transfer calculations, we investigate the
relationship between observable ($L_{IR}$ and $I_{CO}$) and physical
(\sigmastar\ and \sigmagas) variables in these most extreme star-forming
environments.

The next section describes the Eddington-limited star-forming disk model, the
calculation of our model database, and the CO radiative transfer
method. Both
the KS and ES star formation laws predicted by these models are
presented in Section~\ref{sect:laws}. The observational star formation
law (the IR-CO relationship) that is predicted by these models and its
relationship to the physical KS and ES laws is described in
Section~\ref{sect:irco}. A discussion and concluding remarks are
presented in Section~\ref{sect:concl}. Throughout this paper the
\emph{total infrared luminosity} ($L_{\mathrm{TIR}}$) is defined as
the $3$--$1100$~\micron\ luminosity \citep{murph11}, and helium is not accounted for in \sigmagas\ or
$\alpha_{\mathrm{CO}}$. Unless otherwise specified all CO intensities
or luminosities refer to the $J=1-0$ rotational transition.

\section{Calculations}
\label{sect:calc}

\subsection{The Eddington-Limited Star-Forming Disk Model}
\label{sub:model}
The Eddington-limited starburst disk model utilized here was developed
by \citet{tqm05}, and a short summary of the relevant details is
presented below (see also \citealt{todd09}).

Star formation is modeled simply as an energy source at a
location $r$ within a one-dimensional single phase medium that rotates
with angular frequency $\Omega$ in the potential of a
galactic bulge (modeled as an isothermal sphere) with dispersion $\sigma$ and central black hole of mass
$M_{\mathrm{BH}}$ (these two quantities are assumed to be related by
$M_{\mathrm{BH}} = 2\times 10^8
(\sigma/200$~km~s$^{-1})^4$~M$_{\odot}$; e.g.,
\citealt{tre02}). Gas, radiation and turbulent pressure driven by
supernova all combine to provide the vertical support against gravity:
\begin{equation}
\label{eq:pressure}
p_{\mathrm{gas}}+\epsilon\dot{\Sigma}_{\ast}c\left({\tau_{V} \over 2}+\xi\right)=\rho h^2 \Omega^2,
\end{equation}
where $p_{\mathrm{gas}}=\rho kT/m_{\mathrm{p}}$, and $\rho h^2
\Omega^2$ is the total pressure required for hydrostatic balance in
the bulge potential (with
$h$ denoting the gas scale height and $\rho$ indicating the gas
density of the star-forming disk; the self-gravity of the disk is
neglected here). \citet{tqm05} show that in the optically thick limit the
radiation pressure is $\epsilon\dot{\Sigma}_{\ast}c\tau_{V}/2$ where $\epsilon$ is the IMF-dependent efficiency between SFR and luminosity
(i.e., $L=\epsilon \dot{M}_{\ast}c^2$, where $\dot{M}_{\ast}$ is the
local SFR) and $\tau_{V}=\kappa \Sigma_{\mathrm{gas}}/2$ is the vertical optical
depth in the IR ($\kappa(T, \rho)$ is the Rosseland mean opacity of the dusty gas
at temperature $T$ and density $\rho$). The turbulent pressure driven by supernova can also
be written as proportional to the local SFR density,
$\epsilon\xi\dot{\Sigma}_{\ast}c$, where $\xi \approx 1$
\citep{tqm05}. From above, a rms speed $c_{s}$ can be defined using
the total pressure as $c_s=h\Omega$, where it is understood that gas,
radiation and turbulent pressure all contribute to this rms
speed. As described below, the models studied in this paper all have a
radially-averaged $\tau_{V} > 1$ and are thus dominated by radiation
pressure for almost all radii. 

The star-forming disk is assumed to be always just unstable to gravitational
instabilities; that is, Toomre's $Q$ parameter is set equal to
one. Material with gas fraction $f_{\mathrm{gas}}$ is fed onto to the
disk at a radius $r_{\mathrm{out}}$ from the black hole, and, through
a hypothesized global torque (provided by, e.g., a spiral instability
or a bar; \citealt{good03,es04,maj04}), can
slowly accrete towards the center with a radial velocity equal to a
fixed fraction $m$ of the local rms speed \citep{good03}, i.e.,
$\dot{M}=2\pi r \Sigma_{\mathrm{gas}} mc_s$. At any radius $r$,
\citet{tqm05} showed that $\dot{M}$ and $f_{gas}$ can be related through 
\begin{equation}
\label{eq:fgas}
f_{\mathrm{gas}} = \left ( {2^{3/2} \dot{M} G \over Qm\Omega^3 r^3}
\right )^{1/2}.
\end{equation}
As the gas moves through
the disk, it is also producing stars at a rate of
\sigmastar, which, in the optically thick limit, is
determined by requiring radiation pressure in the IR to support the
disk (eq.~\ref{eq:pressure}) as well as $Q=1$:
\begin{equation}
\label{eq:sigmadotstar}
\dot{\Sigma}_{\ast} = {\sqrt{2} f_{\mathrm{gas}} Q \sigma^2 \over
  \epsilon \kappa c r},
\end{equation}
where $r$ is assumed to be large
enough that the black hole potential is negligible. The star formation
slowly consumes the gas as $r$ decreases, reducing both $\dot{M}$ and
$f_{\mathrm{gas}}$, i.e.,
\begin{equation}
\label{eq:mdot}
\dot{M}=\dot{M}_{\mathrm{out}}-\int_{r_{\mathrm{out}}}^r 2\pi r^{\prime}
\dot{\Sigma}_{\ast} dr^{\prime},
\end{equation}
where $\dot{M}_{\mathrm{out}}$ is the mass accretion onto the disk at
$r_{\mathrm{out}}$. 
However, the $Q=1$ assumption causes a rising density in the
inner regions of the disk that substantially increases $\tau_{\mathrm{V}}$ so that
even a moderately increasing \sigmastar\ can still support the disk through
radiation pressure. The solution at radius $r$ is determined by
finding the minimum central gas temperature $T$ and associated
$\kappa$ to ensure vertical
pressure balance, either through the sum of gas and turbulent pressure, or, if
there is enough star formation, radiation pressure. Eventually, either the gas
available for star formation becomes too small to maintain $Q=1$, or
energy release through gas accretion pushes $Q > 1$, and the
calculation ceases.

Substituting eq.~\ref{eq:fgas} into eq.~\ref{eq:sigmadotstar} results
in predictions for both the ES and KS laws in the disks:
\begin{equation}
\label{eq:ESlaw}
\dot{\Sigma}_{\ast}=\left ({2^{3/2} \pi Q G \over \epsilon^2 c^2} {c_s
  \over \kappa^2} \Sigma_{\mathrm{gas}} \Omega \right)^{1/2}
\end{equation}
and, using $Q$ to eliminate $\Omega$ (see \citealt{tqm05}, eq. 3),
\begin{equation}
\label{eq:KSlaw}
\dot{\Sigma}_{\ast}=\left ({2^{1/2} \pi Q G \over \epsilon c
  \kappa}\right )\Sigma_{\mathrm{gas}}.
\end{equation}
These equations have to be viewed carefully because, as seen in
eq.~\ref{eq:mdot}, \sigmagas\ is not an independent variable; its
value at some radius $r$ will depend on \sigmastar\ at larger
radii. As seen in Sect.~\ref{sect:laws}, this property may affect the predicted slopes of the star formation laws. 
In sum, radiation pressure supported star-forming
disks predict a KS relation with $N \approx 1$, and an ES law with $n
\approx 0.5$. Of course, both of these predictions rely on the assumptions
that $Q=1$ and the disk is optically thick in the IR and therefore radiation
pressure supported.

In order to determine how these laws will manifest in
observational surveys of several galaxies, the full disk models need to be calculated over a
wide range of parameters. There are five input parameters for each model: the black
hole mass $M_{\mathrm{BH}}$ (which determines $\sigma$), the angular momentum parameter $m$, the
outer radius $r_{\mathrm{out}}$, the gas fraction at the outer radius
$f_{\mathrm{gas}}$, and a dust-to-gas multiplicative factor for the
opacity to account for the enhanced metallicity observed in the
centers of galaxies \citep[e.g.,][]{mm09}. As in \citet{tqm05}, the \citet{sem03} calculation
of the Rosseland mean opacity for dusty interstellar gas is used for
all models. Calculations are performed for each permutation of
$\log(M_{\mathrm{BH}}/$M$_{\odot})=7,7.5,8,8.5$,
$m=0.0075,0.01,0.025,0.05,0.075,0.1,0.2$, $r_{\mathrm{out}}=50, 100,
150, 200, 250$~pc, $f_{\mathrm{gas}}=0.1,0.5,0.9$ and dust-to-gas
ratios equal to $1\times$, $5\times$ and $10\times$ the local ISM
value, resulting in a suite of 1260 models. While the models are
limited to $r_{\mathrm{out}} \leq 250$~pc, the results on the
star-formation laws do not depend on this size (as is seen for the observed KS and ES laws), and are valid for any
Eddington-limited star-forming region. The range of input gas
fractions is consistent with the observed estimates
\citep[e.g.,][]{tac10,mag12} that show an increase to $\sim$50\% at $z \sim 2$  (the $f_{\mathrm{gas}}$ input to the model drops steadily
through the disk as the star formation uses up the
available gas; eqs.~\ref{eq:fgas} \& \ref{eq:mdot}). We employ a a standard Kroupa/Salpeter IMF with
$\alpha_{\mathrm{IMF}} = -2.35$ \citep{sal55,kw03} between $1$ and
$100$~M$_{\odot}$ and $\alpha_{\mathrm{IMF}} = -1.3$ between $0.1$ and
$1$~M$_{\odot}$ (where the number of stars with masses between $M$ and $M+dM$ is
proportional to $M^{\alpha_{\mathrm{IMF}}}$). This IMF is then input
into a Starburst99 stellar synthesis model \citep{lei99} to compute the bolometric luminosity
of a stellar population with a constant SFR and with the given IMF as a
function of time. The value of $\epsilon=L/\dot{M}c^2$ is then
measured at an age of $10^8$~yr yielding $\epsilon=7.1\times 10^{-4}$
(very similar to value of $6.2\times 10^{-4}$ used by \citealt{os11})\footnote{The assumption of $Q=1$ and the location of the star formation in the
galactic nucleus results in significant gas densities with $\rho$
rising from $\sim 10^{-20}$~g~cm$^{-3}$ ($148$~M$_{\odot}$~pc$^{-3}$)
to $\ga 10^{-14}$~g~cm$^{-3}$ ($1.48\times
10^8$~M$_{\odot}$~pc$^{-3}$). Recently, \citet{krou12} argued that there is a
density and metallicity dependence to the IMF and proposed that the IMF slope
above $\approx 1$~M$_{\odot}$ flattens to one that
depends on density, $\alpha_{\mathrm{IMF}} = -1.86 +
0.43\log(\rho/10^6$M$_{\odot}$~pc$^{-3})$ when $\rho > 9.5\times
10^4$~M$_{\odot}$~pc$^{-3}$. To test the observational consequences of
this IMF, the suite of 1260 models run with the
traditional Kroupa-Salpeter IMF was re-run with a top-heavy IMF
$\alpha_{\mathrm{IMF}}=-1.86$ between $1$ and
$100$~M$_{\odot}$. This IMF yields $\epsilon=1.5\times
10^{-3}$. There were no qualitative differences
in any of the results of this paper between the two IMFs.}. 

\begin{figure}
\includegraphics[width=0.5\textwidth]{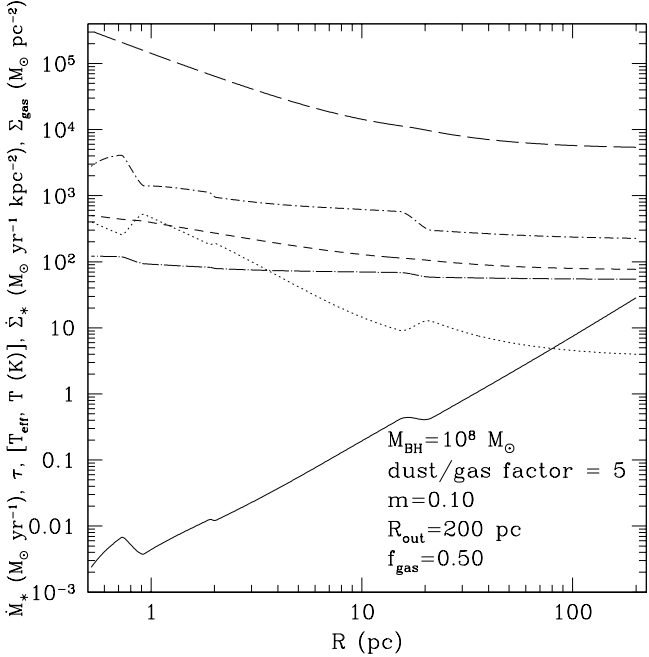}
\caption{\label{fig:model}Several properties of an example Eddington-limited star-forming
  disk calculated using the theory described in
  section~\ref{sub:model} including the SFR (solid line), the vertical
optical depth in the IR $\tau$ (dotted line), the effective
temperature $T_{\mathrm{eff}}$ (dot-long-dashed line), the central
  temperature $T$ (short dashed line), the star formation surface
  density \sigmastar\ (dot-short-dashed line), and the gas surface
  density \sigmagas\ (long dashed line).}
\end{figure}

As described in detail by \citet{tqm05}, roughly two different types of
star-forming disks result from this theoretical description. In about
one-third of the models the gas maintains a
large optical depth over the entire disk, resulting in the temperature
in the inner parsec of the disk exceeding the dust sublimation
temperature and causing a significant burst of star formation in
order to maintain $Q=1$. These parsec-scale bursts could puff up the
disk, and, if they do not use up all the gas, may both fuel and
obscure the central black hole. The observational properties of these
disks have been studied elsewhere \citep{ball08,ab12}, and, as they are
not directly related to the study of general star-forming galaxies,
the description of their star formation laws are deferred to later
paper. The remaining models reach the maximum SFR, SFR$_{\mathrm{max}}$,
at $r_{\mathrm{out}}$ with the SFR declining toward smaller $r$. Many of these
disks are actually gas-pressure dominated or do not extend very far
before \sigmastar\ becomes too small to maintain $Q=1$. As our
interest here is studying radiation pressure dominated star-forming
disks that can extend tens to hundreds of parsecs, we
selected only models with a radially-averaged $\tau_{V} > 1$ and an
inner radius $< 0.04r_{\mathrm{out}}$ for further investigation
(including less extended disks only increases the scatter of the derived relations). This selection criteria yields 132 model star-forming disks. Figure~\ref{fig:model} shows several of the properties of one of
these Eddington-limited star-forming disks. 
The plot clearly shows the implications of fixing $Q=1$: the
increasing \sigmagas\ keeps $\tau_{V}$ large enough so that radiation
pressure dominates over nearly $200$~pc. This fact, in turn, keeps the
disk very warm with temperatures $T > 100$~K and $\dot{\Sigma}_{\ast}
\sim 10^3$~M$_{\odot}$~yr$^{-1}$~kpc$^{-2}$.

Figure~\ref{fig:model} illustrates that the quantities that enter into
the star formation laws (i.e., SFR, \sigmastar\ and \sigmagas) can vary
dramatically over the size of the star-forming disk. As these
variations will be unresolved at most redshifts, we consider both the radial
average and the mass-weighted average of these variables when constructing the
predicted KS and ES laws. For example, the mass-weighted average of
the SFR along the disk, $\langle$SFR$\rangle_{mw}$, is calculated as follows:
\begin{equation}
\label{eq:mwavg}
\langle\mathrm{SFR}\rangle_{mw} = {{\displaystyle{\sum} (2\pi r \Delta r
  \Sigma_{\mathrm{gas}}(r)/f_{\mathrm{gas}}(r))\mathrm{SFR}(r)\ \Delta
  r} \over {\displaystyle{\sum} (2\pi r \Delta r
  \Sigma_{\mathrm{gas}}(r)/f_{\mathrm{gas}}(r))\ \Delta r}},
\end{equation}
where $\Sigma_{\mathrm{gas}}/f_{\mathrm{gas}}$ is the total dynamical
mass surface density. The simple radial average, $\langle$SFR$\rangle$, is computed as in
Eq.~\ref{eq:mwavg} but with the $(2\pi r \Delta r
\Sigma_{\mathrm{gas}}/f_{\mathrm{gas}})$ factors omitted.

Finally, it is important to bear in mind the limitations associated with
this model including the lack of any clumpiness which is
found by hydrodynamical simulations of turbulence supported disks
\citep[e.g.,][]{so12}. In addition, as these disks are
Eddington-limited they will may be subject to Rayleigh-Taylor
instabilities \citep{kt12}, and could also drive outflows and feedback
matter and energy into the galaxy \citep{mqt05,at11}. The effects of outflows on the structure of these disks is not
considered here. Feedback from an accreting black hole is also not treated, so
comparisons between the predictions and data should be limited to
sources without active nuclei. It is expected that the impact of these
limitations will be mostly offset by considering radially averaged
quantities, a wide range of model parameters, and by comparing to
constraints derived from observations of many dozens of individual galaxies. 

\subsection{Infrared Luminosity and SFR}
\label{sub:sfr}
The SFR is one of the most important observational properties of a
galaxy as it gives crucial information on the galaxy's gas content and evolutionary state. Estimating the SFR typically relies on
utilizing one of a
number of relationships between SFR and the luminosity of the galaxy
in a specific wavelength or energy range \citep{ke12}. As most of the starlight
produced in high-$z$ star-forming galaxies is absorbed by dust and
re-emitted in the IR, the SFR-$L_{\mathrm{TIR}}$ relationship
determined by \citet{kenn98b} and \citet{murph11} is one of the most frequently used SFR
estimators, as well as serving as the calibration for SFR estimates at
single IR wavelengths \citep{rieke09,calz10}. As described by
\citet{kenn98b} and \citet{murph11}, the
SFR-$L_{\mathrm{TIR}}$ relationship was calculated using Starburst99
models assuming that 100\% of the bolometric luminosity of a stellar
population with a constant SFR is captured and emitted in the IR. This scenario is also
fulfilled by the optically thick star-forming disk models with
$\langle \tau_V\rangle > 1$, as $\tau_{V}$ is the vertical optical
depth in the IR. Thus, as the SFR is known throughout each disk, it is
interesting to determine how accurately the Murphy \etal-derived SFRs
describe the star-forming properties of the model disks.

Since these disks are largely optically thick at all
radii, the effective temperature $T_{\mathrm{eff}}$ of one side of the
disk surface at radius $r$ is related to half of the total flux produced by
star formation: $\sigma_{\mathrm{SB}}T_{\mathrm{eff}}^4 =
(1/2)\epsilon\dot{\Sigma}_{\ast}c^2$, where $\sigma_{\mathrm{SB}}$ is
the Stefan-Boltzmann constant \citep{tqm05}. Assuming that each annulus of the disk
radiates as a blackbody and is viewed face-on, the IR spectral energy
distribution (SED) of each model is calculated as
\begin{equation}
\lambda L_{\lambda}={2 \pi h c^2 \over \lambda^4}
\int^{r_{out}}_{r_{in}} {2 \pi r dr \over \exp[hc/\lambda
    k_{\mathrm{B}} T_{\mathrm{eff}}(r)]-1},
\label{eq:sed}
\end{equation}
where $r_{\mathrm{in}}$ is the inner radius for each model, $h$ is
Planck's constant and $k_{\mathrm{B}}$ is Boltzmann's
constant. The $L_{\mathrm{TIR}}$ for each model is then computed by
integrating the SED from $3$--$1100$~\micron. This luminosity can then
be compared to various calculations of the disk SFR, such as
$\langle$SFR$\rangle$, SFR$_{\mathrm{max}}$, and
$\langle$SFR$\rangle_{mw}$.

\begin{figure}
\includegraphics[width=0.5\textwidth]{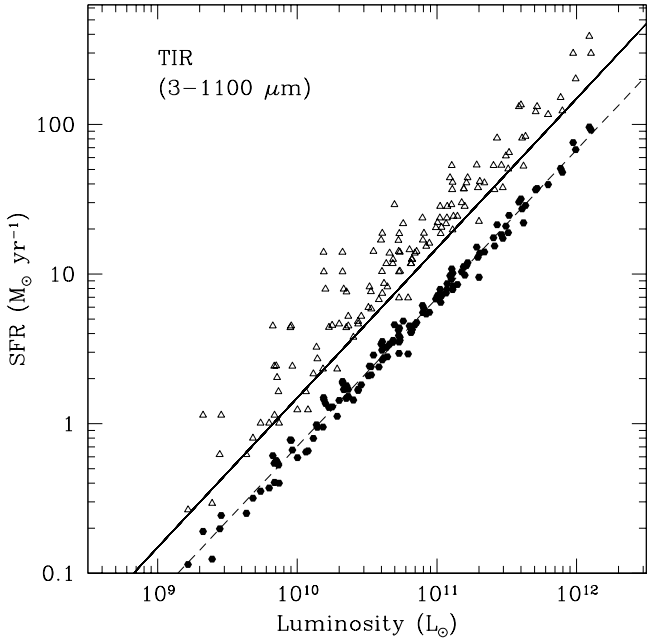}
\caption{\label{fig:sfrlir}Radially averaged (filled circles) and maximum (open
  triangles) SFRs from Eddington-limited star-forming
  disks plotted as a function of the infrared luminosity (computed from the
  spectrum derived from eq.~\ref{eq:sed}). The short dashed line
  denotes the least-squares fit to the data while the solid line plots the SFR calibration published by \citet{murph11}.}
\end{figure}

Figure~\ref{fig:sfrlir} plots the \citet{murph11} calibration
($\log($SFR$/$M$_{\odot}$~yr$^{-1})=\log(L_{\mathrm{TIR}}/\mathrm{erg\ s}^{-1})-43.41$) as the solid line, SFR$_{\mathrm{max}}$
versus $L_{\mathrm{TIR}}$ (open triangles), and $\langle$SFR$\rangle$
against $L_{\mathrm{TIR}}$ (solid points). The SFR$_{\mathrm{max}}$,
$\langle$SFR$\rangle$ and $L_{\mathrm{TIR}}$ are derived directly from
the models as described above.
Evidently, the \citet{murph11} calormetric relation provides a coarse
approximation to the maximum SFR of each disk. This is a natural consequence of the
assumption of a single SFR that underlies the original computation of
the SFR-$L_{\mathrm{TIR}}$ relationship. The maximum SFR in the disk
provides the largest contribution to the luminosity, so the estimated
SFR from the $L_{\mathrm{TIR}}$ relationships will be closest to the maximum
rate. However, as shown in Fig.~\ref{fig:sfrlir}, a better (with $\sim
10\times$ smaller scatter) SFR-$L_{\mathrm{TIR}}$ relationship is found with the
$\langle$SFR$\rangle$ which is, on average, $2.2\pm 0.3$ smaller than
the values predicted by the \citet{murph11} calibration at the same luminosity. A least squares fit to the
$\langle$SFR$\rangle$-$L_{\mathrm{TIR}}$ relationship (dashed line)
yields $\log(\langle$SFR$\rangle/$M$_{\odot}$~yr$^{-1})=(0.991\pm0.008)\log(L_{\mathrm{TIR}}/$L$_{\odot})-(10.07\pm
0.09)$. Thus, the $\langle$SFR$\rangle$ is used throughout
the paper to quantify the SFR for any particular model.

\subsection{Computing the CO Line Emission}
\label{sub:co}
The CO intensity for multiple rotational lines is computed for each
model using the one-dimensional version of the molecular line
radiative transfer code \textsc{ratran} \citep{hv00}. \citet{ab12}
previously used \textsc{ratran} to study the CO spectral line energy
distributions (SLEDs) of eighteen star-forming disk models (nine with
a pc-scale burst and nine without), and a complete description of the
calculation procedure is described in that paper. Below is a summary
of the key parameters and assumptions that enter into the calculation.

The temperature and density of the star-forming disk models are computed on a logarithmic radial grid
which is input into \textsc{ratran}. 
As mentioned above the Eddington-limited star-forming disks are
uniformly very dense, so that the gas is expected to be entirely
molecular as long as $T \la 1000$~K \citep[e.g.,][]{pel06}. Thus, the molecular
fraction of the gas is set to unity for all radii, and the CO
abundance is set to be $10^{-4}$ of the gas number density
\citep{kemp06}. Since the star-forming gas is under high pressure and
typically has densities $> 10^{4}$~cm$^{-3}$, gas
and dust will be well mixed, and the molecular kinetic temperature and dust temperatures are set to the
central temperature $T$ \citep{gold01}. Dust opacity is included in the radiative
transfer calculation with a temperature dependent broken power-law
opacity (see \citealt{ab12}), and is scaled as appropriate for each
model's dust-to-gas factor. The turbulent line width at each radius is
derived from the local rms speed $c_s$ (i.e., the Doppler broadening
parameter is $b = (2\sqrt{\ln(2)})^{-1} c_s$), and the radial velocity is
$mc_s$, as defined above. Once \textsc{ratran} completes the radiative
transfer calculation for a disk model, the Miriad software
package \citep{sa95} is used to remove the appropriate dust emission from
each spectral line and integrate over velocity to produce the
velocity-integrated intensity $I_{\mathrm{CO}}$ in K~km~s$^{-1}$. Using this method, CO line
intensities for the $J=$1--0, 2--1, 3--2, 4--3, 5--4, 6--5, 7--6, 8--7
and 9--8 transitions are predicted for each Eddington-limited
star-forming disk model.

\section{The Star Formation Laws}
\label{sect:laws}

\subsection{The Kennicutt-Schmidt Relation}
\label{sub:ks}
\begin{figure*}
\begin{center}
\includegraphics[angle=-90,width=1.0\textwidth]{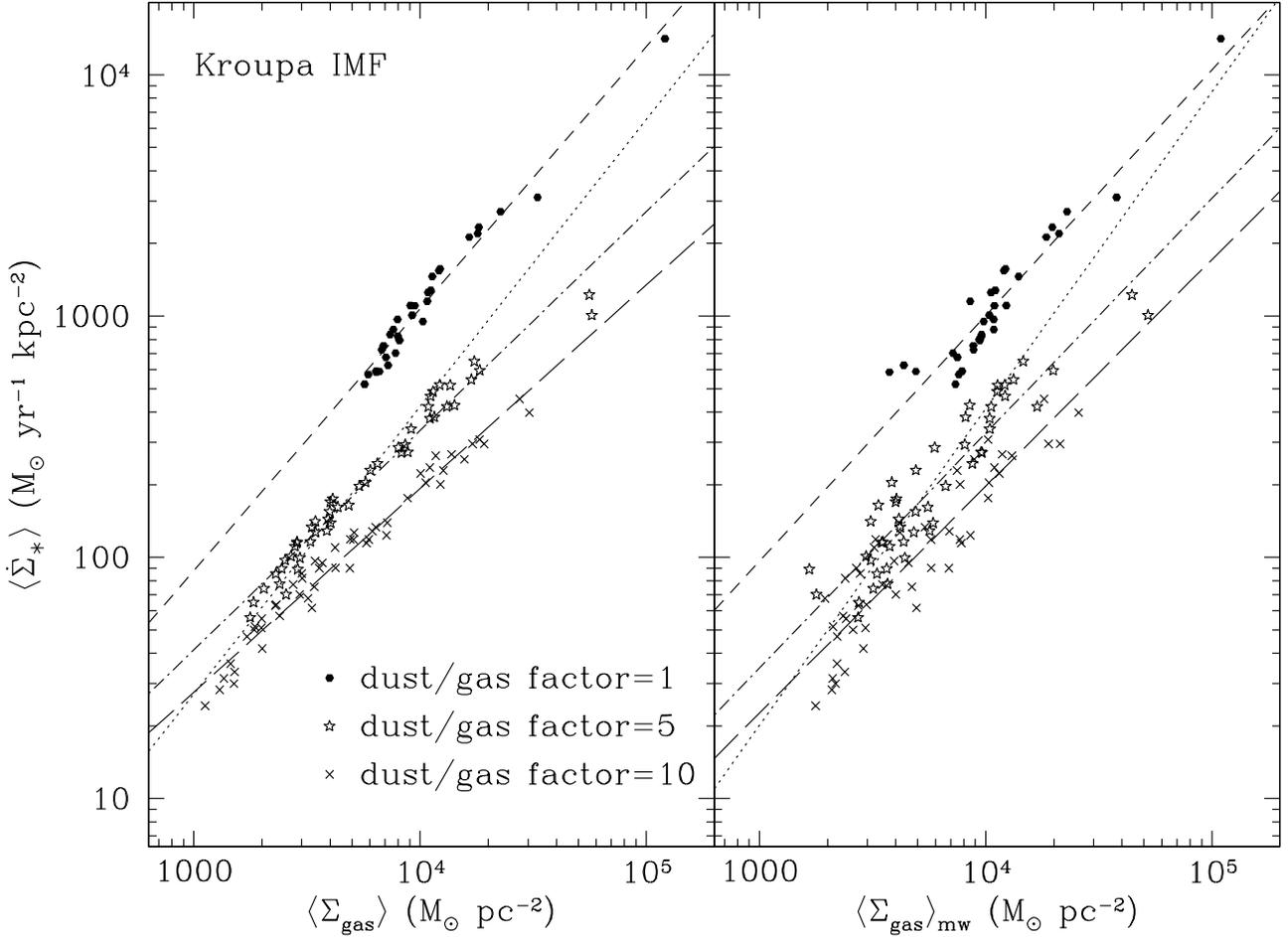}
\caption{\label{fig:kslaw}Kennicutt-Schmidt relations, defined either as
  $\langle\dot{\Sigma}_{\ast} \rangle$
  vs. $\langle\Sigma_{\mathrm{gas}}\rangle$ (left panel) or as $\langle\dot{\Sigma}_{\ast} \rangle$
  vs. $\langle\Sigma_{\mathrm{gas}}\rangle_{mw}$ (right panel), predicted by the
  Eddington-limited starburst disk models. The slope of the relations are slightly dependent on the dustiness of the gas with $N=1.1$,
  $0.91$ and $0.84$ when plotted versus $\langle\Sigma_{\mathrm{gas}}\rangle$ and a dust-to-gas enhancement of $1\times$,
  $5\times$ and $10\times$, respectively. When
  $\langle\dot{\Sigma}_{\ast} \rangle$ is plotted against
  $\langle\Sigma_{\mathrm{gas}}\rangle_{mw}$ the slopes are
  $N=1.0$, $0.97$ and $0.94$ as the dustiness of the gas increases. The
  dotted line in each panel plots the relation when all models are
  considered, irrespective of the dust-to-gas factor. In that case,
  $N=1.2$ (left panel) and $1.3$ (right panel), very similar to
  the KS relations measured from less intense star-forming regions at low
  and high redshift \citep[e.g.,][]{ke12}. 
}
\end{center}
\end{figure*}

Turning now to the star formation laws predicted by the
Eddington-limited star-forming disks, Figure~\ref{fig:kslaw} plots the
KS law in two ways: $\langle$\sigmastar$\rangle$ versus
$\langle$\sigmagas$\rangle$ (left panel) and $\langle$\sigmastar$\rangle$ against
$\langle$\sigmagas$\rangle_{mw}$ (right panel).
The theoretical expectation is for
$\dot{\Sigma}_{\ast} \propto \Sigma_{\mathrm{gas}}/\kappa$
(Eq.~\ref{eq:KSlaw}), and, indeed, we find that the KS laws,
calculated from these radial averages, have slopes close to unity and
are seperated vertically by the dust-to-gas enhancement factor (that
increases $\kappa$). The scatter in the KS relations is smallest when
plotted against $\langle$\sigmagas$\rangle$ because the same spatial
averaging is being performed on both quantities. However, due to the
dependence of \sigmagas\ on \sigmastar\ (see eq.~\ref{eq:mdot} and
surrounding discussion), the slopes of the KS laws in the left-hand
panel can drift away from the expectation of $N=1$. This effect is
enhanced for larger dust-to-gas enhancement factors because a larger
$\kappa$ reduces \sigmastar\ and thus \sigmagas\ does not decrease as
fast with radius. The spatial averaging therefore results in a larger
\sigmagas\ which then flattens the slope of the relations. By plotting
the quanitites with different spatial averaging, the
right-hand panel breaks the dependence of \sigmagas\ on
\sigmastar\ and shows that indeed the KS laws follow the theoretical
expectation with $N\approx 1$. 

A random selection of galaxies will likely exhibit a wide range of
dust-to-gas ratios that may be difficult to observationally
seperate. The dotted lines in Fig.~\ref{fig:kslaw} plot the KS
relations found when including all the model points, regardless of the
dust-to-gas factor. The resulting slopes are $N=1.2$ and $1.3$ for the 
$\langle$\sigmagas$\rangle$ and $\langle$\sigmagas$\rangle_{mw}$
panels, respectively. These values are very similar to the observed KS
laws \citep[e.g.,][]{ke12}, and indicate that the common practice of measuring the KS
relation by compiling a large collection of heterogeneous data may be
hiding important clues of the physics of star-forming disks. Future
investigations of the star formation laws may benefit from the
analysis of a smaller sample of galaxies with well known properties.

\subsection{The Elmegreen-Silk Relation}
\label{sub:es}
\begin{figure*}
\begin{center}
\includegraphics[angle=-90,width=1.0\textwidth]{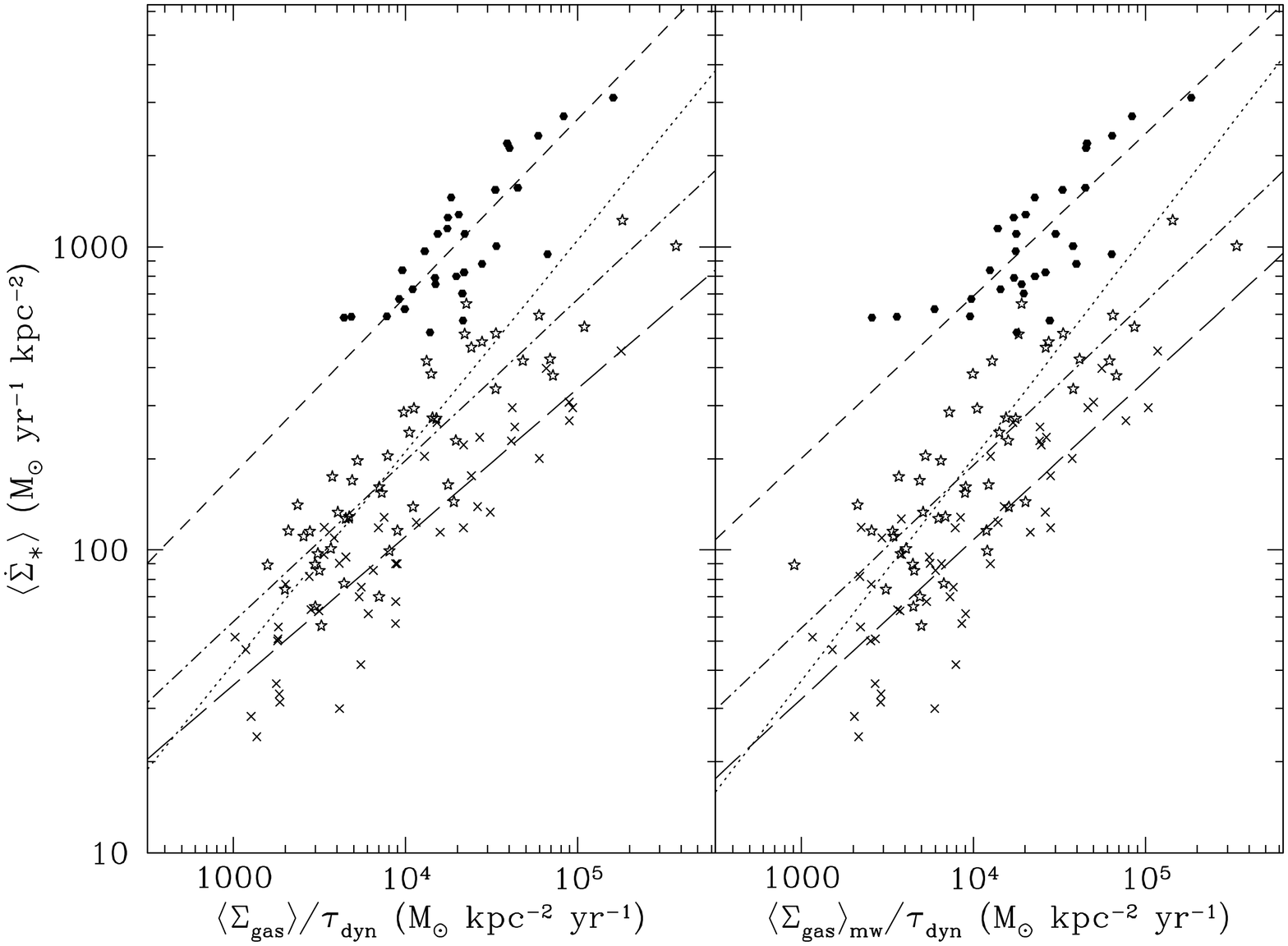}
\caption{\label{fig:eslaw}Elmegreen-Silk relations, defined either as
  $\langle\dot{\Sigma}_{\ast} \rangle$
  vs. $\langle\Sigma_{\mathrm{gas}}\rangle/\tau_\mathrm{dyn}$ (left panel) or as $\langle\dot{\Sigma}_{\ast} \rangle$
  vs. $\langle\Sigma_{\mathrm{gas}}\rangle_{mw}/\tau_\mathrm{dyn}$ (right panel), predicted by the
  Eddington-limited starburst disk models. The dynamical time,
  $\tau_{\mathrm{dyn}}$, is defined as the inverse of the orbital
  frequency at $r_{\mathrm{out}}$, i.e.,
  $\tau_{\mathrm{dyn}}=1/\Omega(r_{\mathrm{out}})$. As
  for the KS laws (Fig.~\ref{fig:kslaw}), the models are divided based on the
  multiplicitive dust-to-gas factor with the same symbol styles as the
  previous figure. When $\langle\dot{\Sigma}_{\ast} \rangle$ is
  plotted versus
  $\langle\Sigma_{\mathrm{gas}}\rangle/\tau_\mathrm{dyn}$ (left panel)
  the power-law slopes are equal to $n=0.59$, $0.53$ and $0.49$ for a
  dust-to-gas enhancement of $1\times$, $5\times$ and $10\times$,
  respectively. If the $\langle\dot{\Sigma}_{\ast} \rangle$ is instead
  plotted against
  $\langle\Sigma_{\mathrm{gas}}\rangle_{mw}/\tau_\mathrm{dyn}$ (right panel) the
  slopes are $n=0.54$, $0.54$ and $0.53$. The dotted line plots the
  least-squares fit to the all the points and has a slope of $n=0.70$
  (left panel) and $0.73$ (right panel).}
\end{center}
\end{figure*}

The ES law predicted from the radiation pressure dominated star-forming
disk models are shown in Fig.~\ref{fig:eslaw},
with the same symbols and line styles as the previous KS law
figure. Following the current observational practice the orbital
frequency is set to the inverse of the dynamical timescale at
$r_{\mathrm{out}}$, i.e., $\tau_{\mathrm{dyn}}=1/\Omega(r_{\mathrm{out}})$.
The theoretical expectation for the ES law is
\sigmastar$\propto$(\sigmagas$\Omega/\kappa^2)^{0.5}$ (eq.~\ref{eq:ESlaw}), which is in
good agreement with the relations seen in Fig.~\ref{fig:eslaw}. There
is a flattening of the slope with the dust-to-gas enhancement
factor that is reduced in the
$\langle\Sigma_{\mathrm{gas}}\rangle_{mw}/\tau_\mathrm{dyn}$ plot. The
explanation for this effect is the same as with the KS relations
discussed above. However, it is clear from a comparison of
figs.~\ref{fig:kslaw} and~\ref{fig:eslaw} that the ES law has
significantly more scatter than the KS relations (about a $3\times$
larger rms in the $\langle\Sigma_{\mathrm{gas}}\rangle$ plots; $\sim
50$\% larger rms in the $\langle\Sigma_{\mathrm{gas}}\rangle_{mw}$
panels). This increased scatter is likely due to dividing a radial
average by the orbital time at one radius, as well as a contribution
due to variations in the rms speed in each model
(eq.~\ref{eq:ESlaw}). It is notable that even when all the models are
fit with a powerlaw, irrespective of the dust-to-gas factor, the
measured slopes ($n=0.70$ and $0.73$) are still significantly
sub-linear. Thus, measurements of a sub-linear ES-law in high-$z$ star-forming
galaxies may provide compelling evidence for a high-density Eddington
limited environment.

\section{The Infrared--CO Relationship and the Observed Star Formation
Laws}
\label{sect:irco}
The previous section described the theoretical expectations for the
star formation laws governing the radiation pressure supported
disks. To test the theory, these laws must be observationally inferred
from data of a sample of galaxies, which, at high-$z$, are typically
limited to photometric measurements such as an IR luminosity and a
molecular line intensity. The challenge is then to turn these measured
quantities into accurate estimates of \sigmastar\ and \sigmagas\ for
each galaxy. In this section, we make use of our predicted IR
luminosities and $I_{\mathrm{CO}}$ for each disk model\footnote{We
  work with the velocity integrated intensity $I_{\mathrm{CO}}$ instead of $L^{\prime}_{\mathrm{CO}}$ to
  more closely match the observables expected for high-$z$ galaxies.} to test how
well the observationally derived star formation laws recover the
theoretical laws presented in Sect.~\ref{sect:laws}. 

\begin{figure}
\begin{center}
\includegraphics[width=0.5\textwidth]{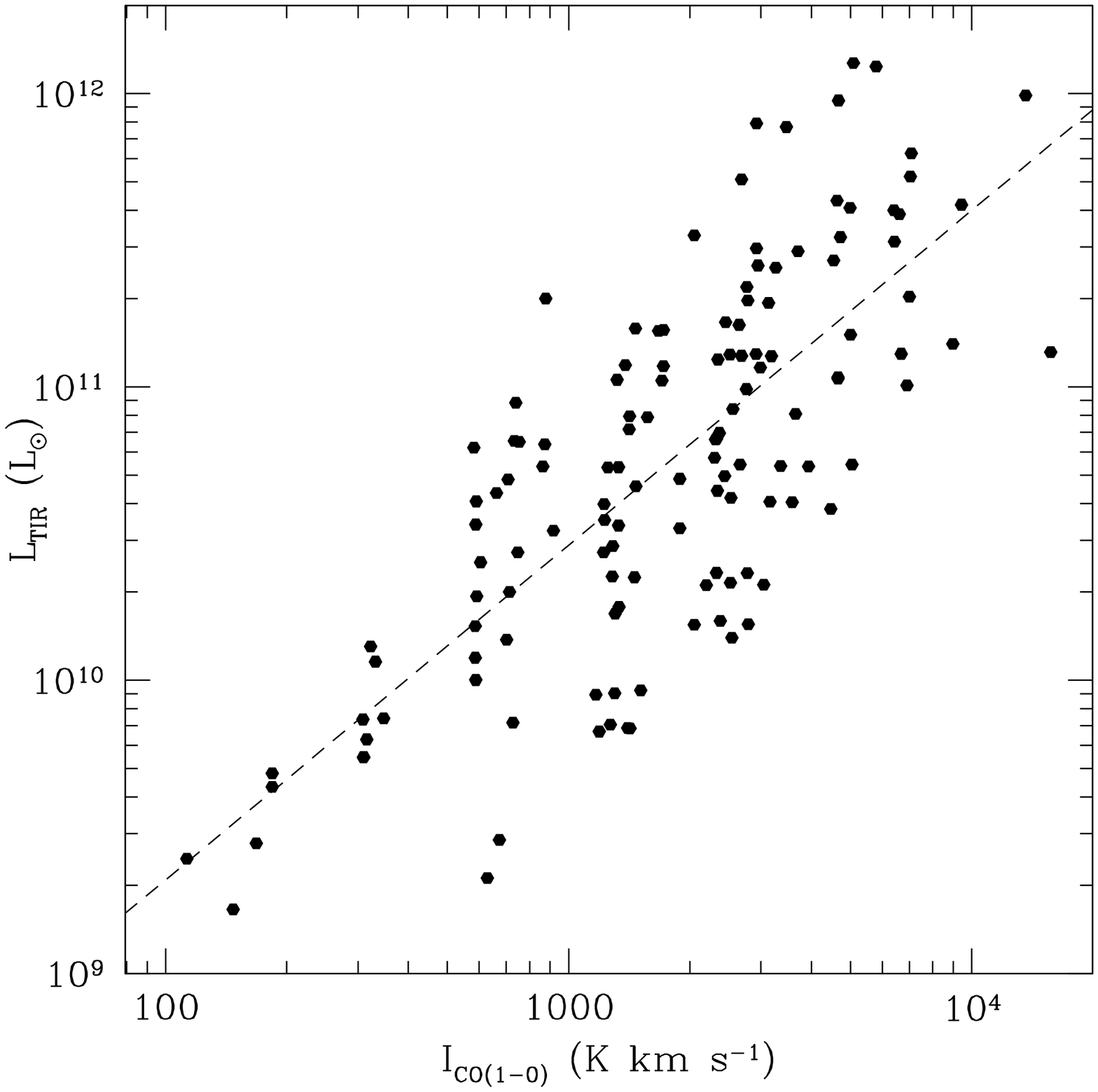}
\caption{\label{fig:lirico}The $L_{\mathrm{TIR}}$ and $I_{CO}$ relationship obtained
  from the Eddington-limited star-forming disk models. The dashed line shows the least-squares fit to the model data
  with a slope of $1.1\pm 0.1$. If this plot is interpreted as an
  observational analogue of the KS law, then the slope is in
  reasonable agreement with the theoretical expectation of $1.2$ or $1.3$
  (Fig.~\ref{fig:kslaw}). However, this quasi-linear relation is simply
a result of the molecular line and dust emission tracing the same gas
population, and is not an accurate estimate of the underlying KS law.}
\end{center}
\end{figure}

Since an IR luminosity is correlated with the SFR (Sect.~\ref{sub:sfr})
and the molecular line intensity is connected \emph{in some manner} to
the total gas content of galaxy, a common zeroth-order measurement of the KS law is made by plotting
the observed IR luminosity against a molecular line luminosity or
intensity (Figure~\ref{fig:lirico}).
The CO $J=1-0$ transition is used as the molecular line tracer in this
plot because it is one of the most commonly measured lines; however, 
the densities and temperatures of these disks are large enough that
all the low $J$ CO lines are fully thermalized and thus have approximately
equal luminosities (see \citet{ab12} for example CO SLEDs from the
radiation pressure dominated disks). As a result, neither the slope
nor the scatter in the $L_{\mathrm{TIR}}$-$I_{\mathrm{CO}}$
relationship depend significantly on the CO transition. 

Interestingly, the slope of the $L_{\mathrm{TIR}}$-$I_{\mathrm{CO}}$
relationship ($1.1\pm 0.1$) is very similar to the one predicted
from the theoretical KS law ($N=1.2$ or $1.3$ when including all
models; Fig.~\ref{fig:kslaw}). However, this is purely a
coincidence, because in these warm, dense and well-mixed disks both the
dust and molecular gas emitters are thermalized, optically thick and represent the
same amount of disk mass. Thus, a quasi-linear relationship
between the $L_{\mathrm{TIR}}$ (indicating the dust mass) and
$I_{\mathrm{CO}}$ (indicating the gas mass) is expected in this
scenario \citep{kt07}. We conclude that $L_{\mathrm{TIR}}$-$I_{\mathrm{CO}}$
plots are poor estimates of the star formation laws of radiation
pressure dominated disks.

The next step is to use the observables to estimate the physical
quantities that enter into the star formation laws, beginning with
\sigmastar. This quantity is reasonably straightforward to calculate,
as one can simply translate the $L_{\mathrm{TIR}}$ into a SFR through one of the
well known calibrations and then divide by the observed area of the
disk to obtain the star formation surface density.
\begin{figure}
\begin{center}
\includegraphics[width=0.5\textwidth]{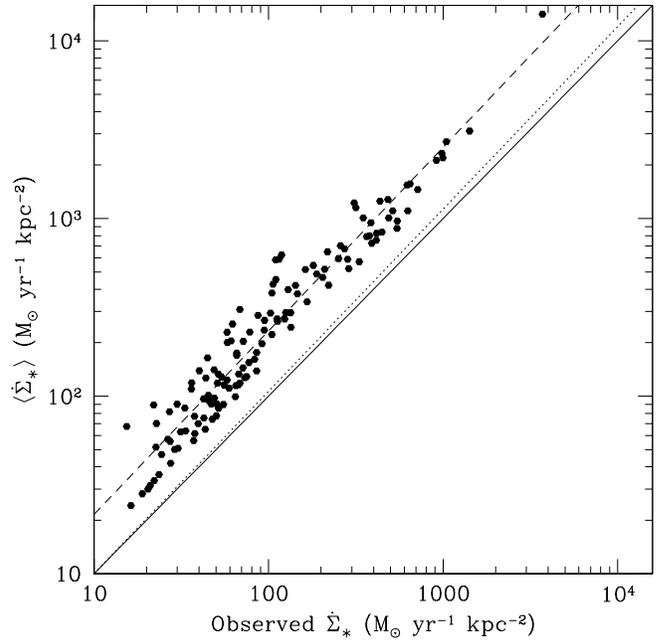}
\caption{\label{fig:sigmadotstars}A comparison between the $\langle\dot{\Sigma}_{\ast}
  \rangle$ predicted by the Eddington-limited starburst disks and the one calculated from the `observed'
  $L_{\mathrm{TIR}}$, a $L_{\mathrm{TIR}}$-SFR correlation, and the
  inner and outer radii of the disk. The solid line plots the
  one-to-one correspondence between the two quantities, the dotted
  line is the relationship found when the \citet{murph11}
  $L_{\mathrm{TIR}}$-SFR calibration is used to derive $\langle\dot{\Sigma}_{\ast}
  \rangle$, while the data points and the dashed line result from the
  $L_{\mathrm{TIR}}$-$\langle$SFR$\rangle$ correlation described in Sect.~\ref{sub:sfr}. This figure indicates that, due to the
  spatial variation of the star formation rate, any simple estimate of
$\langle\dot{\Sigma}_{\ast} \rangle$ will be inaccurate by a factor of
$\la 2$.}
\end{center}
\end{figure}
Figure~\ref{fig:sigmadotstars} illustrates how well that procedure
recovers the actual $\langle\dot{\Sigma}_{\ast}\rangle$ of the model
star-forming disks by plotting $\langle\dot{\Sigma}_{\ast}\rangle$
versus the values obtained from $L_{\mathrm{TIR}}$, the relationship from Sect.~\ref{sub:sfr}, and the model inner and
outer radii. The `observed' \sigmastar\ is highly correlated with
$\langle\dot{\Sigma}_{\ast}\rangle$, and the least squares fit to the
data (dashed line) has a slope of $1.04\pm 0.02$, but is $\approx
2\times$ smaller than the one-to-one relationship. The dotted line is
the least-squares fit to the model data if the \citet{murph11}
$L_{\mathrm{TIR}}$-SFR relationship is used and results in a
\sigmastar\ that is approximately correct at low
$\langle\dot{\Sigma}_{\ast}\rangle$, but becomes progressively too
small at large $\langle\dot{\Sigma}_{\ast}\rangle$. These offsets from the
actual $\langle\dot{\Sigma}_{\ast}\rangle$ is a result of
approximating the average \sigmastar\ with
($\langle$SFR$\rangle$)/area or (SFR$_{\mathrm{max}}$/area), but are relatively minor. Most
importantly, there is a good linear correlation between
$\langle\dot{\Sigma}_{\ast}\rangle$ and the observational estimate,
which indicates that the translation from $L_{TIR}$ results in a
reasonable estimate of $\langle\dot{\Sigma}_{\ast}\rangle$.

\begin{figure}
\begin{center}
\includegraphics[width=0.5\textwidth]{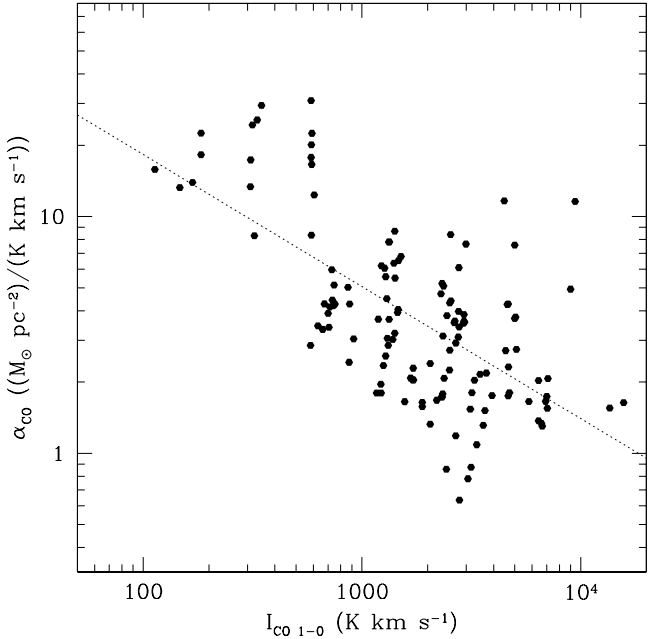}
\caption{\label{fig:alphaico}The relationship between $\alpha_{\mathrm{CO}}=\langle\Sigma_{\mathrm{gas}}\rangle_{mw}/I_{\mathrm{CO}}$ and
  $I_{\mathrm{CO}}$ obtained from the radiation pressure dominated
  star-forming disks. The dotted line is the
  least-squares fit to the model data with a slope of $-0.56\pm
  0.06$. As $I_{\mathrm{CO}}$ is correlated with the
  $\langle$SFR$\rangle$ (Fig.~\ref{fig:lirico}), this plot also shows
  that $\alpha_{\mathrm{CO}}$ is anti-correlated with the $\langle$SFR$\rangle$.}
\end{center}
\end{figure}

As described in Sect.~\ref{sect:intro} the conversion factor
$\alpha_{\mathrm{CO}}$ between $I_{\mathrm{CO}}$ and \sigmagas\ is not
well determined except for a small number of galactic star-forming regions and some
local ULIRGs \citep[e.g.,][]{ds98,svb05,nar12}. In our case, since we have both $I_{\mathrm{CO}}$
and $\langle\Sigma_{\mathrm{gas}}\rangle_{mw}$ for every model, we can
calculate an unique value of $\alpha_{\mathrm{CO}}$ for each
star-forming disk (calculating $\alpha_{\mathrm{CO}}$ with
$\langle\Sigma_{\mathrm{gas}}\rangle$ instead of
$\langle\Sigma_{\mathrm{gas}}\rangle_{mw}$ results in more scatter, so
we focus on those computed with the mass-weighted average). Figure~\ref{fig:alphaico} plots the relationship
between the resulting $\alpha_{\mathrm{CO}}$ and $I_{\mathrm{CO}}$ and
indeed finds a reasonable anti-correlation ($R=-0.64$, where $R$ is
the linear correlation coefficient) with a slope of
$-0.56\pm 0.06$. The median value of $\alpha_{\mathrm{CO}}$ is $3.56$.
As the CO intensity is correlated with the SFR
(Fig.~\ref{fig:lirico}), this plot also indicates that
$\alpha_{\mathrm{CO}}$ is anti-correlated with the SFR. The slope of the $\alpha_{\mathrm{CO}}$-$I_{\mathrm{CO}}$
anti-correlation is steeper than the one of $-0.32$ found by \citet{nar12} from
hydrodynamical simulations of galaxies. Similarly, the amplitude
of the dotted line in Fig.~\ref{fig:alphaico} is $\ga 2\times$ larger
than the \citet{nar12} relation. As the physical scenario and calculation of the
CO intensity performed by \citet{nar12} differs significantly from our setup
the similarity in the results is striking. The relatively high gas temperatures found in
the Eddington-limited star forming disks will tend to drive
$\alpha_{\mathrm{CO}}$ lower \citep{magnelli12,nar12}, which may explain the
steeper slope in Fig.~\ref{fig:alphaico}. The larger amplitude may be caused by the lack of
clumpiness in our star-forming medium which, when combined with the optically-thick
nature of the CO emission, would result in the CO line intensity being
a worse tracer of the total gas mass than in a clumpy medium where the
CO intensity is directly related to a large number of smaller
clouds. However, \citet{pap12} recently argued that the centers of ULIRGs (which may be radiation pressure dominated; \citealt{tqm05}) are unlikely
to be comprised of individual molecular clouds and that the small values
of $\alpha_{\mathrm{CO}}$ inferred from these environments may be erroneously
low, in agreement with the results presented in
Fig.~\ref{fig:alphaico}. 

\begin{figure}
\begin{center}
\includegraphics[width=0.5\textwidth]{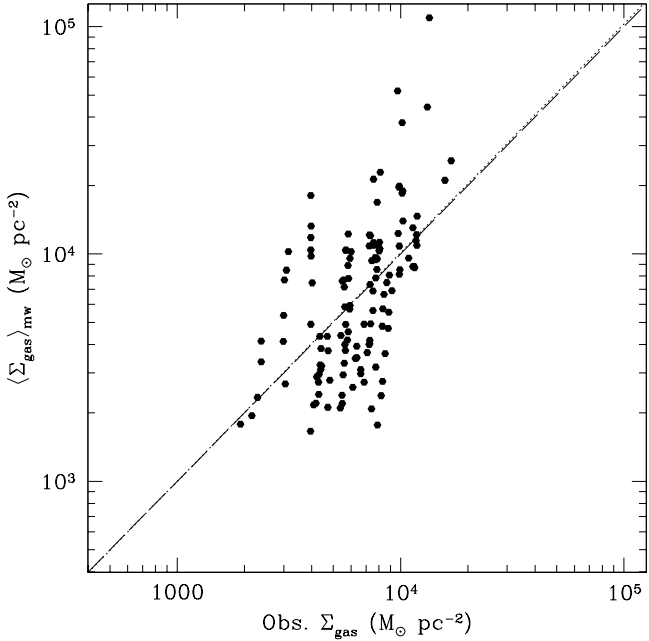}
\caption{\label{fig:sigmagas}This figure compares the value of
  $\langle\Sigma_{\mathrm{gas}}\rangle_{mw}$ derived directly from the
star-forming disk models with those calculated from the `observed'
$I_{\mathrm{CO}}$ and the $\alpha_{\mathrm{CO}}$-$I_{\mathrm{CO}}$
relationship shown in Fig.~\ref{fig:alphaico}. The long-dashed line is
the one-to-one relationship and the dotted line is the least-squares
fit to the data. While the two values agree on average, there is only a
weak correlation ($r=0.55$) between the
$\langle\Sigma_{\mathrm{gas}}\rangle_{mw}$ calculated from the model
and the one derived from $I_{\mathrm{CO}}$.}
\end{center}
\end{figure}

It is apparent from Fig.~\ref{fig:alphaico} that there is no single
value of $\alpha_{\mathrm{CO}}$ for a given $I_{\mathrm{CO}}$;
indeed, the scatter in $\alpha_{\mathrm{CO}}$ at one
$I_{\mathrm{CO}}$ can reach a factor of ten. Using the best-fit
$\alpha_{\mathrm{CO}}$-$I_{\mathrm{CO}}$ relation will than propagate
 this large scatter through to the estimated \sigmagas\ (as seen in Fig.~\ref{fig:sigmagas}).
The values of \sigmagas\ derived from $\alpha_{\mathrm{CO}}$ are only
weakly correlated ($R=0.55$) with the
$\langle\Sigma_{\mathrm{gas}}\rangle_{mw}$ calculated directly from
the models. This poor relationship between $I_{\mathrm{CO}}$ and
\sigmagas\ will substantially complicate any attempt to recover the
star formation laws from the observational quantities. 

\begin{figure*}
\begin{center}
\includegraphics[angle=-90,width=1.0\textwidth]{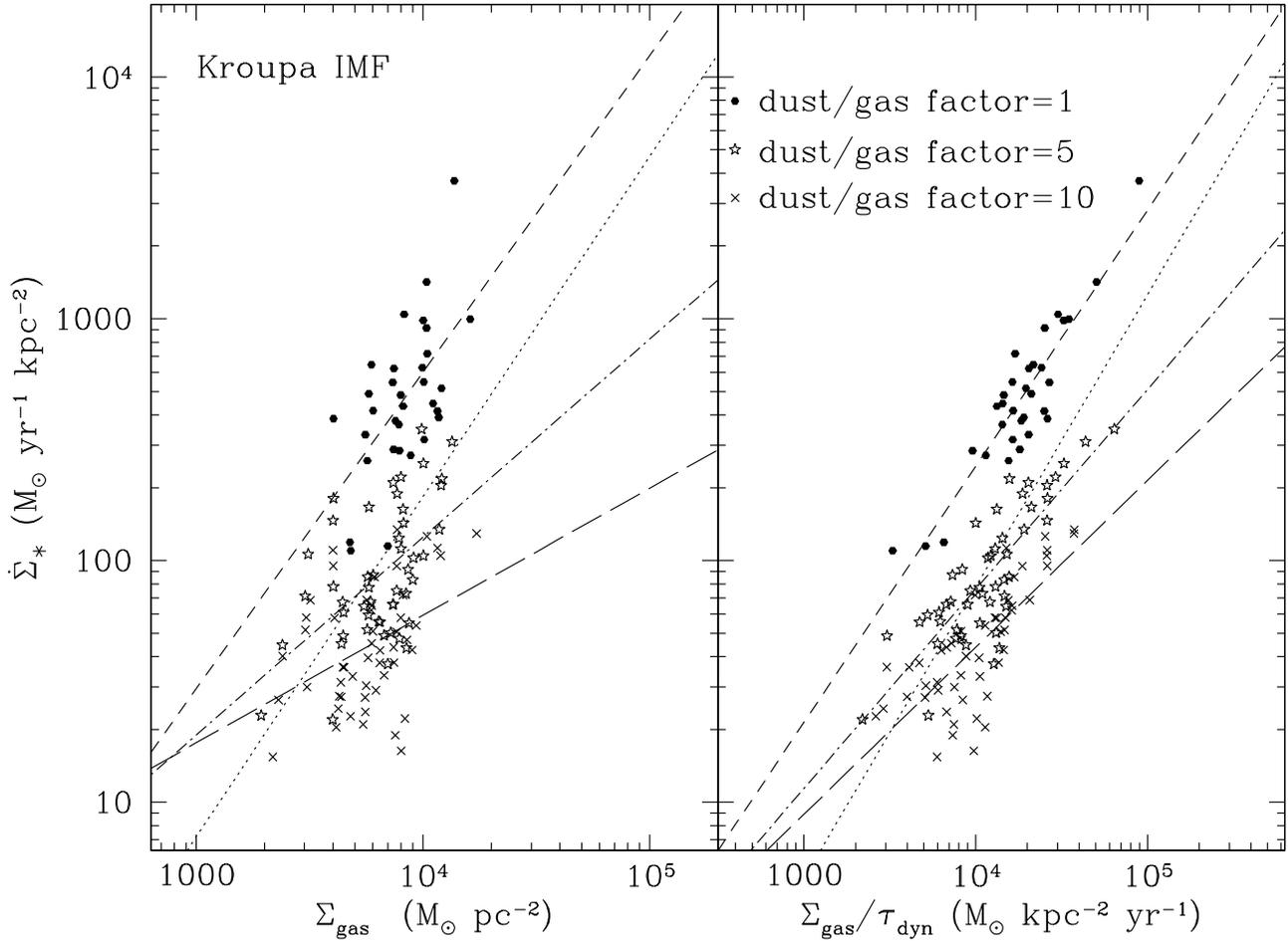}
\caption{\label{fig:sflaws}The KS (left panel) and ES (right-panel) laws of radiation pressure supported star-forming
  disks derived from the $L_{\mathrm{TIR}}$-$I_{\mathrm{CO}}$
  relationship (Fig.~\ref{fig:lirico}) as described in the
  text. As in the plots of the theoretically expected star formation
  laws (Figs.~\ref{fig:kslaw} and~\ref{fig:eslaw}), the models have
  been separated by the value of the dust-to-gas enhancement
  factor. The slope of the relations are dependent on the
  dustiness of the gas with the KS relations yielding $N=1.3$,
  $0.8$ and $0.5$ for a dust-to-gas enhancement of $1\times$,
  $5\times$ and $10\times$, respectively. For the ES relations the slopes are
  $n=1.1$, $0.8$ and $0.7$ as the dustiness of the gas increases. The
  dotted line in each panel plots the relation when all models are
  considered, irrespective of the dust-to-gas factor. In that case,
  $N=1.4\pm 0.2$ (KS law) and $1.2\pm 0.1$ (ES law). Overall, the laws
derived from observations are poor representations of the actual
star formation relations that govern the disk.}
\end{center}
\end{figure*}

Finally, Figure~\ref{fig:sflaws} presents the KS and ES laws of the
radiation pressure supported star-forming disks derived
solely from the observational quantities $L_{\mathrm{TIR}}$ and
$I_{\mathrm{CO}}$. To guide the analysis and comparison with the
theoretical expectations (figs.~\ref{fig:kslaw} and~\ref{fig:eslaw}),
the models are separated by their dust-to-gas enhancement factor,
although this may be impractical for a real observational
sample. Recall that the \sigmagas\ computed from $I_{\mathrm{CO}}$ is
an estimate of $\langle\Sigma_{\mathrm{gas}}\rangle_{mw}$, so
Fig.~\ref{fig:sflaws} should be compared to the right-hand panels of Figs.~\ref{fig:kslaw} and~\ref{fig:eslaw}.
As expected, the scatter of the `observed' KS laws (rms$\approx 0.2$)
is larger than the ones found from the theoretical
relations (rms$\approx 0.1$). More striking are the differences in
the correlation coefficient and slopes. The observationally derived KS
laws have slopes of $N=1.3\pm 0.3$, $0.8\pm 0.2$, and $0.5\pm 0.2$ and
correlation coefficients of $R=0.6$, $0.5$ and $0.4$ as the
dust-to-gas enhancement factor increases. In contrast the
theoretically expected slopes are $0.94$--$1.0$ and have correlation
coefficients of $>0.9$. Collecting all the models together
yields a KS slope of $N=1.4\pm 0.2$ (with $R=0.5$), surprisingly close
to the canonical value, but, of course, yielding an entirely
misleading view of the underlying physics. Thus, the KS law governing the
Eddington-limited star-forming disks is nearly unrecognizable due to
the very poor conversion between $I_{\mathrm{CO}}$ and
\sigmagas. Our data indicate that the use of a higher density CO
transition (such as CO~$J=3-2$) would not improve the conversion as
this line is thermalized in these very warm disks. A high-$J$
transition from a large density tracing molecule, such as HCN, may
provide the best probe of Eddington-limited disks (although this needs to be
confirmed by subsequent computations).

In contrast to the KS law, the `observational' ES laws (right panel of
Fig.~\ref{fig:sflaws}) have similar correlation coefficients
($R\approx 0.8$) and about the same amount of scatter (rms$\approx
0.2$) as the theoretically expected relations (right panel of
Fig.~\ref{fig:eslaw}). However, the slopes of the different ES laws
are all $n>0.7$, inconsistent with the expected slope of $n=0.5$. The
slope for the entire ensemble of models is $n=1.2\pm0.1$, which, much like the
KS law, leads to an erroneous description of the correct ES law. The
steep slopes are a result of the uncertain
$\alpha_{\mathrm{CO}}$-$I_{\mathrm{CO}}$ relation. If, for example, a
constant $\alpha_{\mathrm{CO}}=3.56$, was used to estimate \sigmagas,
then the slopes flatten to $n=0.4$-$0.7$ for the individual dust-to-gas
factors, and $n=0.8\pm0.1$ when including all the models. These
results are consistent with the theoretical ES law from
Sect.~\ref{sub:es}. Applying this constant $\alpha_{\mathrm{CO}}$
to the observed KS law results in a slope that is much too flat
($N=0.2$--$0.6$). The slopes of the observed ES and KS relations are
therefore very sensitive to the assumptions placed on
$\alpha_{\mathrm{CO}}$; however, the low rms and large correlation
coefficients found in the observed ES law are robust to the
assumptions on $\alpha_{\mathrm{CO}}$, indicating that, out of these
two relationships, the ES law may provide the most promising means to
investigate SF laws at high redshift.

\section{Discussion \& Conclusions}
\label{sect:concl}
The purpose of this paper is to, one, elucidate the KS and
ES laws of a Eddington-limited star-forming disk (including the impact of
spatial averaging), and, two, to determine how well the theoretically
expected star formation laws can be recovered by the standard
conversions from the observables ($L_{IR}$ and $I_{\mathrm{CO}}$). 
To explore the effects of spatial averaging and how
these disks might appear to telescopes, 1260 star-forming
disk models, spanning a wide range of size and mass scales, were run
to produce a bank of test data. From this model suite, 132 of the largest
and most optically thick disks were selected to best describe the
properties of Eddington-limited starbursts that may be embedded in the
centers of the gas-rich star-forming galaxies. 

The analytical model of a radiation pressure supported star-forming disk
developed by \citet{tqm05} predicts a KS law with $N\approx 1$, and an ES law
with $n \approx 0.5$, both significantly flatter than the canonical values
($N\approx 1.5$ and $n\approx 1$). An important aspect of these star formation laws is the
dependence on the opacity, or, equivalently, the dustiness of the star-forming
gas. Indeed, the star formation process in a radiation supported
disk strongly depends on the dust opacity \citep{tqm05}, and, when
spatially averaged, can even impact the slopes of the observed star
formation laws (Figs.~\ref{fig:kslaw} and~\ref{fig:eslaw}). Another important effect of the gas opacity is the vertical
separation of the star formation laws based on the dust-to-gas
enhancement factor (Figs.~\ref{fig:kslaw} and~\ref{fig:eslaw}). A
heterogeneous sample of Eddington-limited star-forming disks may then
appear to have a significantly steeper slope than what is governing
the disks and would lead to the wrong conclusion about the
underlying star formation relations.

To study how these disks might appear observationally, molecular line
radiative transfer calculations were performed and CO
line velocity-integrated intensities, $I_{\mathrm{CO}}$, were derived
for all 132 star-forming disk models. Combining these results with the infrared luminosity,
$L_{\mathrm{TIR}}$, calculated from the SED of each model, allowed a
quantitative test of deriving the star formation laws from
observational data. The calculated $L_{\mathrm{TIR}}$ and $I_{\mathrm{CO}}$ are approximately
linearly correlated with each other (albeit with a substantial
scatter of rms$=0.4$; Fig.~\ref{fig:lirico}), as might have been
expected from the theoretical KS law. However, in the
Eddington-limited star-forming disks, the gas density and temperature
are always above the critical density and excitation temperature of
the first several CO rotational transitions. These lines are therefore
completely thermalized and are tracing the same gas as the infrared
emission, resulting in a near-linear relationship \citep[e.g.,][]{kt07,nar08}. Indeed, the
densities and temperatures of these disks are so large that a linear
relationship is also expected for the low $J$ transitions of molecules with higher critical
densities (such as HCN or HCO$^+$), as seen in some ULIRGs
\citep[e.g.,][]{gb12}. While such a linear relationship may indicate that the
star-forming region could be radiation-pressure dominated, it does not
give any information on the underlying KS law
\citep{kt07}. Observations of high-$J$ transitions from these
molecules may be needed to determine the KS and ES laws from the observed IR
luminosity and molecular line intensity. 

If the $L_{\mathrm{TIR}}$-$I_{\mathrm{CO}}$ relation cannot be trusted
as a measurement of the star formation laws of Eddington-limited
disks, then these quantities must be translated to \sigmastar\ and \sigmagas. Figure~\ref{fig:sigmadotstars} shows that the use of a
$L_{\mathrm{TIR}}$-SFR conversion and the outer
radius of the disk results in a $\langle\dot{\Sigma}_{\ast}\rangle$ that
is within a factor of 2 of the correct value. On the other hand, $\alpha_{\mathrm{CO}}$, the
phenomenological conversion factor between $I_{\mathrm{CO}}$ and
\sigmagas, exhibits significant scatter (about an order of magnitude
for a given $I_{\mathrm{CO}}$; Fig.~\ref{fig:alphaico}) and yields a
very weak correlation between the estimated \sigmagas\ and the true
$\langle\Sigma_{\mathrm{gas}}\rangle_{mw}$ (Fig.~\ref{fig:sigmagas}). 
This poor conversion from $I_{\mathrm{CO}}$ to \sigmagas\ causes the
resulting KS laws (Fig.~\ref{fig:sflaws} (left)) to have only a weak
correlation between \sigmastar\ and \sigmagas. The slopes of this KS
relation vary strongly with the dustiness of the gas, reaching as
low as $N=0.5\pm 0.2$ for a dust-to-gas factor of 10. The slope of
$N=1.4\pm 0.2$ found from including all the models is consistent with
the canonical slope, which perfectly illustrates how the
observed KS law can distort and mislead the interpretation of star
formation laws.

The situation is improved somewhat when considering the ES law,
calculated by dividing the $I_{\mathrm{CO}}$-derived \sigmagas\ by
the orbital time at $r_{\mathrm{out}}$ (Fig.~\ref{fig:sflaws}
(right)). In this case, the observed ES laws retain the large
correlation coefficients expected from the theoretical predictions,
but, as with the KS law, the slopes are significantly steeper ($n >
0.7$) than what the theory predicts ($n\approx 0.5$). Similar to the KS law,
the slope found from the ensemble of models ($n=1.2\pm 0.1$) is close
to the canonical linear slope, leading to yet another erroneous
conclusion about the underlying star formation law.

These experiments all point to the same problem: the
$\alpha_{\mathrm{CO}}$ conversion factor is too blunt and inaccurate
an instrument to study the star formation laws in radiation pressure
dominated disks. In many ways this is unsurprising, as the
$\alpha_{\mathrm{CO}}$ factor has only been shown to accurately trace
\sigmagas\ in local Galactic molecular clouds up to \sigmagas$\approx
100$~M$_{\odot}$~pc$^{-2}$ \citep{ke12}. The Eddington-limited disks are at
such high surface densities that the medium may not be accurately described as a
series of molecular clouds, but rather a continuous
structure with a near unity molecular fraction. In their study of the
CO SLEDs from these disks, \citet{ab12} noted that in many ways the
thermalized molecular emission from these disks acts like a stellar spectrum, with a SLED shape that
depends on excitation temperature and a normalization that depends on
the internal energy generation, i.e., star formation. Therefore, a
large, massive disk may produce a similar $I_{\mathrm{CO}}$ as a
smaller, less massive disk that has a larger $\langle$SFR$\rangle$
(due to, e.g., a larger opacity). The CO emission is largely
immune to to the total gas mass in each disk, and therefore the
$\alpha_{\mathrm{CO}}$ is a poor tracer of \sigmagas. We do find that
$\alpha_{\mathrm{CO}}$ is inversely related to $I_{\mathrm{CO}}$ and
$\langle$SFR$\rangle$, in agreement with other authors \citep{nar12}, but
with an amplitude $\approx 2\times$ larger. This may be an effect of
the high density and uniform molecular medium of these disks \citep{pap12},
but this result should be checked with more realistic two- or three-dimensional
simulations.

Given the difficulties in using $\alpha_{\mathrm{CO}}$, how can we make
progress in studying star formation laws, especially when
\sigmagas$\ga 
10^4$~M$_{\odot}$~pc$^{-2}$? Clearly the traditional method of CO
rotational lines will not be particularly useful in this regime, so
observations should focus on molecular species with higher critical
densities, such as HCN. Recent results using HCN observations of ULIRGs that reach
into this high-\sigmagas\ range do give KS and ES slopes consistent
with the theoretical expectation for Eddington-limited disks,
suggesting that radiation pressure dominated regions may be playing a role in these sources
\citep{gb12}. Unfortunately, the conversion factor of HCN and other
higher density tracers is as uncertain as $\alpha_{\mathrm{CO}}$. Observational campaigns focusing
on measuring these conversion factors over as wide a range of
\sigmagas\ as possible will be necessary in order to confidently apply
them for ULIRGS and at high redshift where large gas surface densities will be more
common. Interestingly, the ES law appears to be more robust to changes
in opacity and maintains a strong correlation even with the use of
$\alpha_{\mathrm{CO}}$ (although the slope is still incorrect). This
result implies that, at least for the Eddington-limited star-forming
disks studied here, measurements of the ES law will be more useful in
determining the underlying star formation relations than the KS
law. Finally, the star formation laws of radiation pressure dominated
disks are strongly dependent on the gas opacity, causing a vertical
separation in the KS and ES plots. Fitting a line through the models
without controlling for the differences in the dust-to-gas ratio
results in a much steeper slope than the underlying star formation
law. Thus, future observational work at these values of
\sigmagas\ would benefit from analyzing samples of galaxies with
metallicity or dust-to-gas ratio estimates in order to control for
this potentially misleading effect.

In summary, the KS and ES star formation laws for radiation pressure
supported star-forming disks have flatter slopes ($N\approx1$ and $n\approx0.5$) than what is expected
for star formation at \sigmagas$\la 10^4$~M$_{\odot}$~pc$^{-2}$. These
slopes are relatively robust to spatial averaging over the disks, but
the strong dependence on the opacity causes the relations to be
vertically offset based on the dust-to-gas ratio. Thus, an erroneously large slope for the KS and ES
laws could result if the opacity dependence is not
recognized. Attempting to recover these star formation laws from the
predicted $L_{\mathrm{TIR}}$ and $I_{\mathrm{CO}}$ failed because of
the poor translation from $I_{\mathrm{CO}}$ to \sigmagas\ at these
high densities and temperatures. The `observed' KS and ES laws have
slopes that differ significantly from the theoretically expected
relations, even when controlling for the gas opacity. Progress in
identifying and studying the star formation laws at these values of
\sigmagas\ can be made by moving to well calibrated high critical
density molecular tracer, selecting a sample of galaxies with
metallicity estimates, and focusing on the ES relation, as its slope
seems less affected by variations in metallicity and spatial averaging.

\acknowledgments
The authors thank D.\ Narayanan for comments on a draft of the
paper. This work was supported in part by NSF award AST 1008067 to DRB.

{}

\end{document}